\input{aipcheck}

\documentclass[
    ,final            
  ]
  {aipproc}

\layoutstyle{6x9}

\newcommand{\nn}{\nonumber\\}
\newcommand{\ep}{\epsilon}

\newcommand{\be}{\begin{equation}}
\newcommand{\bw}{\begin{widetext}}
\newcommand{\ew}{\end{widetext}}
\newcommand{\bse}{\begin{subequation}}
\newcommand{\ese}{\end{subequation}}
\newcommand{\ee}{\end{equation}} 
\newcommand{\eei}{\end{eqnarray}\indent\indent}
\newcommand{\bc}{\begin{center}}
\newcommand{\ec}{\end{center}}
\newcommand{\ber}{\begin{eqnarray}}
\newcommand{\eer}{\end{eqnarray}}
\newcommand{\ba}{\begin{array}}
\newcommand{\ea}{\end{array}}
\newcommand{\bal}{\begin{align}}
\newcommand{\eal}{\end{align}}

\newcommand{\hs}{\,-\,}

\newcommand{\sfrac}[2]{{\textstyle{#1\over#2}}}
\def\case#1/#2{\textstyle\frac{#1}{#2} }
\newcommand{\nb}{\nabla}
\newcommand{\D}{\tl\nb}

\newcommand{\tl}{\tilde}
\newcommand{\bver}{\begin{verbatim}}
\newcommand{\ever}{\end{verbatim}}

\begin{document}

\title{Simultaneous expansion and rotation of  shear-free universes in modified gravity}


\author{Amare Abebe}{
  address={ Astrophysics, Cosmology and Gravity Centre (ACGC),
University of Cape Town, Rondebosch, 7701, South Africa}
}

\author{Rituparno Goswami}{
  address={ Astrophysics, Cosmology and Gravity Centre (ACGC),
University of Cape Town, Rondebosch, 7701, South Africa}
}

\author{Peter K.S. Dunsby}{
  address={ Astrophysics, Cosmology and Gravity Centre (ACGC),
University of Cape Town, Rondebosch, 7701, South Africa}
  ,altaddress={ South African Astronomical Observatory, Observatory, 7925, South Africa} 
}
\begin{abstract}
We show in a fully covariant way that, there exist a class of $f(R)$ models for which a shear-free, almost FLRW universe can expand and rotate at the same time . 
\end{abstract}
\maketitle
\section{Introduction}
The action of a generalized fourth-order gravity is given by (for more details, see \cite{Abebe} and references therein):
\be
{\cal A}= \sfrac12 \int d^4x\sqrt{-g}\left[f(R)+2{\cal L}_m\right]\;,
\label{action}
\ee
where ${\cal L}_m$ represents the matter contribution,
and the generalized field equations read
 \be
 G_{ab}=\tl T^{m}_{ab}+T^{R}_{ab}\equiv T_{ab}\;,
 \ee
where 
 \be
\tl T^{m}_{ab}=\frac{T^{m}_{ab}}{f'}\;,~~~~
T^{R}_{ab}=\frac{1}{f'}\left[\sfrac{1}{2}(f-Rf')g_{ab}+\nb_{b}\nb_{a}f'-g_{ab}\nb_{c}
\nb^{c}f' \right]\,.
\ee
Here we have defined $f'\equiv df(R)/dR$ and  $T^{m}_{ab} = \mu^{m}u_{a}u_{b} + p^{m}h_{ab}+ q^{m}_{a}u_{b}+ q^{m}_{b}u_{a}+\pi^{m}_{ab}$,

where  $\mu^{m}$, $p^m$, $q^{m}$ and $\pi^m_{ab}$ denote the standard matter density, 
pressure, heat flux and anisotropic stress respectively.

The {\it total} thermodynamics of the matter-curvature composite is then given by
\be
\mu\equiv\frac{\mu^{m}}{f'}+\mu^{R}\;,~~~\;p\equiv\frac{p^{m}}{f'}+p^{R}\;,~~~
q_{a}\equiv \frac{q^{m}_{a}}{f'}+q^{R}_{a}\;,~~~\;\pi_{ab}\equiv\frac{\pi^{m}_{ab}}{f'}+\pi^{R}_{ab}\;,
\ee
where $\mu^{R}$, etc. are thermodynamical quantities of the curvature fluid defined in the next section.
The covariant derivative of a timelike vector $u^a$ can  be decomposed into basic parts as
\be
\nb_au_b=-A_au_b+\sfrac13h_{ab}\Theta+\sigma_{ab}+\ep_{a b c}\omega^c,
\ee
where $A_a=\dot{u}_a$ is the acceleration, $\Theta=\tl\nb_au^a$ is the expansion, 
$\sigma_{ab}=\tl\nb_{\langle a}u_{b \rangle}$ is the shear tensor and $\omega^{a}=\ep^{a b c}\tl\nb_bu_c$ 
is the vorticity vector. For the  Weyl curvature tensor one has
\be
E_{ab}=C_{abcd}u^cu^d=E_{\langle ab\rangle}\;,~H_{ab}=\sfrac12\ep_{acd}C^{cd}_{be}u^e=H_{\langle ab\rangle}\;,
\ee
giving a covariant description of {\it tidal forces} and {\it gravitational radiation} respectively.
\section{Linearized Field Equations }
 We consider  the background to be Friedmann-Lem\u{a}itre-Robertson-Walker (FLRW), where the 
Hubble scale sets the characteristic scale of the perturbations. In the perturbed spacetime the 
standard matter is considered to be a perfect fluid  with the energy momentum tensor given by:
\be
T^m_{ab}=(\mu^m+p^m)u_au_b+p^mg_{ab}\;.
\label{EMT}
\ee
with
$p^m=w\mu^m$ and the heat flux ($q^m_a$)  and the anisotropic stress 
($\pi^m_{a b}$) vanishing in the perturbed spacetime.  In addition, since we consider 
shear-free perturbations, the shear tensor $\sigma_{a b}$ vanishes identically.

For the {\it curvature fluid} the linearized thermodynamic quantities are given by
\ber
&&\mu^{R}=\frac{1}{f'}\left[\sfrac{1}{2}(Rf'-f)-\Theta f'' \dot{R}+ f''\tilde{\nabla}^{2}R \right]\;,\nn
&&p^{R}=\frac{1}{f'}\left[\sfrac{1}{2}(f-Rf')+f''\ddot{R}+f'''\dot{R}^{2}+\sfrac{2}{3}\left( \Theta f''\dot{R}-f''\tilde{\nabla}^{2}R \right) \right]\;,\nn
&&q^{R}_{a}=-\frac{1}{f'}\left[f'''\dot{R}\tilde{\nabla}_{a}R +f''\tilde{\nabla}_{a}\dot{R}-\sfrac{1}{3}f''\Theta \tilde{\nabla}_{a}R \right],~~~\pi^{R}_{ab}=\frac{1}{f'}f''\tilde{\nabla}_{\langle a}\tilde{\nabla}_{b\rangle}R\;.
\eer
With the conditions above, the propagation and constraint equations can be given by
\be
\dot{\Theta}-\tl\nb_aA^a=-\sfrac13 \Theta^2-\sfrac12(\mu+3p)\;,
\label{R1}
\ee
\be 
(\omega^{\langle a \rangle})^{.}-\sfrac{1}{2}\ep^{abc}\tl\nb_bA_c=-\sfrac23\Theta\omega^a\;,
\label{R3}
\ee
\be
\dot{E^{\langle a b\rangle}}-\ep^{cd\langle a}\tl\nb_cH^{\rangle b}_d=-\Theta E^{ab}-
\sfrac{1}{2}\dot{\pi}^{ab}_{R}
-\sfrac{1}{2}\tl\nb^{\langle a}q^{b\rangle}_{R}-\sfrac{1}{6}\Theta\pi^{ab}_{R}\;,
\label{B1}
\ee
\be
\dot{H^{\langle ab \rangle}}+\ep^{cd\langle a}\tl\nb_cE^{\rangle b}_d=-\Theta H^{ab}+
\sfrac{1}{2}\ep^{cd\langle a}\tl\nb_c\pi^{\rangle b}_{d~R}\;,
\label{B2}
\ee
\be\label{B4}
\dot{\mu}_{m}=-(\mu_{m}+p_{m})\Theta,
\ee
\be
\dot{\mu}+\tl\nb^{a}q^{R}_{a}=-(\mu+p)\Theta;
\ee
\be
(C_0)^{ a b}:=E^{a b}-\tl\nb^{\langle a}A^{b \rangle}-\sfrac{1}{2}\pi_{R}^{ab}=0\;,
\label{R2}
\ee
\be
(C_1)^a:=\tl\nb^a\Theta-\sfrac{3}{2}\ep^{abc}\tl\nb_b\omega_c-\sfrac{3}{2}q^{a}_{R}=0\;,
\label{R4}
\ee
\be 
(C_2):=\tl\nb^a\omega_a=0\;,
\label{R5}
\ee
\be
(C_3)^{ a b}:=H^{a b}+\tl\nb^{\langle a}\omega^{b \rangle}=0\;,
\label{R6}
\ee
\be
(C_4)^a:=\tl\nb^ap_{m} +(\mu_{m}+p_{m}) A^a=0\;,
\label{B3}
\ee
\be
(C_5)^a:=\tl\nb_bE^{a b}+\sfrac{1}{2}\tl\nb_{b}\pi^{ab}_{R}-\sfrac13\tl\nb^a\mu+
\sfrac13\Theta q^{a}_{R}=0\;,
\label{B5}
\ee
\be
(C_6)^a:=\tl\nb_bH^{a b}+(\mu+p)\omega^a+\sfrac12\ep^{abc}\tl\nb_{b}q_{c}^{R}=0\;.
\label{B6}
\ee
The conditions 
$\sigma_{a b}=0$  and $q^a_{m}=0$ give the two new constraints $(C_0)^{a b}$ and $(C_4)^a$ 
respectively. 
Substituting $(C_0)_{b d}$ 
into $(C_5)_b$ and using $(C_4)_b$ we obtain
 the 
constraint
\be
\frac{w}{w+1}\tl\nb^d\tl\nb_{\langle b}\tl\nb_{d \rangle}\phi+\sfrac{1}{3}\tl\nb_b\mu-\tl\nb^{d}\pi^{R}_{bd}-\sfrac{1}{3}\Theta q^{R}_{b}=0,
\label{newcons}
\ee
where $\phi\equiv \ln{\mu_{m}}$.
 To check the spatial consistency of the 
above constraint on any initial hypersurface we take the curl of (\ref{newcons}) to obtain
\be\label{conricc}
\omega^{a}\left[ \left(\frac{w\Theta}{3}+\frac{\dot{R}f''}{3f'}\right)\tl{R}
+\frac{2(1+w)\mu_{m}\Theta}{3f'}\right]+\left(\frac{\dot{R}f''}{f'}+w\Theta\right)\D^{2}\omega^{a}=0\;,
\ee
where $\tl{R}=2\left(\mu-\frac13\Theta^2\right)$.
Now defining the 
expansion, acceleration, jerk and snap parameters  by the following relations
\be
\Theta=3\frac{\dot a}{a}\;, ~~~~\; ~~~q=-\frac{\ddot{a}a}{\dot{a}^{2}},~~~j=\frac{a^{2}}{\dot{a}^{3}}\frac{d^{3}a}{dt^{3}}\;,~~~~\;s=\frac{a^{3}}{\dot{a}^{4}}\frac{d^{4}a}{dt^{4}}\;,
\ee
and using
\ber
&&\dot{\Theta}=-\frac{1}{3}\Theta^{2}(1+q)\;,~~~\dot{q}=-\frac{1}{3}\Theta\left(j-q-2q^{2}\right)\;,~~~\ddot{\Theta}=\frac{1}{9}\Theta^{3}\left(2+3q+j\right)\;,\nn
&&\dot{j}=\frac{1}{3}\Theta\left(s+2j+3qj\right)\;,~~~\ddot{q}=-\frac{1}{9}\Theta^{2}\left[s+2j-3q^{2}+6qj-6q^{3}\right]\;,~~~\dot{R}=\frac{2}{3}\Theta Q\;,\nn
\eer
where
\be
Q=\frac{1}{3}\Theta^{2}(j-q-2)+\tl{R},~~~~~~~~\dot{Q}=\sfrac{1}{9}\Theta\left[(4+5q+j+jq+s)\Theta^{2}+6\tl{R}\right]\;,
\label{dop}
\ee
we can rewrite (\ref{conricc}) as
\be\label{conric}
\sfrac{2}{3}\Theta\bigg\{\omega^{a}\left[ \left(\frac{w}{2}+\frac{f''}{3f'}Q\right)\tl{R}+\frac{(1+w)\mu_{m}}{f'}\right]+\left[\frac{f''}{f'}Q+\frac{3w}{2}\right]\D^{2}\omega^{a}\bigg\}=0\;.
\ee
Spatial consistency requires the vanishing of  either $\Theta$ or the terms in the curly brackets. For temporal consistency differentiate (\ref{conric}) w.r.t. time to get
\be\label{conricci}
\Theta\omega^{a}\bigg\{\left[\frac{(1-w)P}{3}\tl{R}+\frac{(1+w)}{f'}\frac{(3w+5)f'+4f''Q}{6f'}\mu_{m}\right]+\frac{Z}{P}\left[(\frac{1+w}{f'})\mu_{m}\right]\bigg\}=0\;,
\ee
where
\be 
Z=\frac{2}{3}\left(\frac{f'''}{f'}-(\frac{f''}{f'})^{2}\right)Q^{2}+\frac{f''}{9f'}\left((4+5q+j+jq+s)\Theta^{2}+6\tl{R}\right).
\ee
It follows that for the new constraints to be spatially  and temporally consistent we must have either $\Theta\omega^a=0$ or
the expression in the curly brackets must vanish.  It is easy to see from (\ref{conricci}) that if the 3-curvature vanishes, then $\Theta\omega^{a}=0$ for vacuum universes ($\mu_m=0$). This implies that {\em a shear-free, spatially flat vacuum universe in any $f(R)$ theory can rotate and expand  simultaneously in the linearized regime.} 

In the non\hs vacuum case,  there  exists at least one non-trivial case which does violate the Ellis condition. For a flat Milne universe,  the Friedmann equation is given by
\be
-R^2\frac{d^{2}f(R)}{dR^{2}}+\frac{f(R)}{2}-\frac{\mu_{0}}{a(R)^{3(1+w)}}=0\;,
\ee
and has the following general solution:
\be\label{sol1}
f(R)=C_{1}R^{\frac{1+\sqrt{3}}{2}}+C_{2}R^{\frac{1-\sqrt{3}}{2}}-
\frac{4\mu_{0}}{1+12w+9w^{2}}R^{\frac{3(1+w)}{2}}.
\ee
Considering the particular solution ($C_{1}=0=C_{2}$), 
and comparing it with Eqn. (\ref{conricci}), for the corresponding flat Milne universe in 
$R^{n}$ gravity, we obtain \be\label{sol2}
\frac{(1+w)\mu_{m}}{6f'}\left[3w+9-4n\right]=0.
\ee
Comparing solutions (\ref{sol2}) and the particular solution of 
(\ref{sol1}) (with $n=3(1+w)/2$) we find that $w=1$ if $\mu_{m}\ne0$. In other words, 
{\em for a stiff fluid in $R^3$ gravity, there exists a flat Milne-universe solution 
which can rotate and expand simultaneously at the level of linearized perturbation theory.}
\section{Discussion and Conclusion}
In this work we showed that if the 3-curvature vanishes, then  the result of \cite{GFR} can always be avoided for vacuum universes. We also demonstrated there is at least one physically realistic non-vacuum case in which both rotation and expansion are simultaneously possible. This suggests that there are situations where linearized fourth-order gravity shares properties with Newtonian theory not valid in General Relativity. 
\begin{theacknowledgments}
The authors thank the organisers of the Spanish Relativity Meeting (ERE2011) and the National Research Foundation (South Africa) for financial support. \end{theacknowledgments}

\bibliographystyle{aipproc}   

\bibliography{sample}

\IfFileExists{\jobname.bbl}{}
 {\typeout{}
  \typeout{******************************************}
  \typeout{** Please run "bibtex \jobname" to optain}
  \typeout{** the bibliography and then re-run LaTeX}
  \typeout{** twice to fix the references!}
  \typeout{******************************************}
  \typeout{}
 }

\end{document}